\documentclass[aps,prl,preprint,showpacs,superscriptaddress,floatfix]{revtex4}
\usepackage[latin1]{inputenc}
\usepackage{graphicx}
\usepackage[colorlinks=true,citecolor=blue,linkcolor=blue]{hyperref}
\usepackage{amssymb,amsmath}
\usepackage{color}

\begin{document}

\title{Fractional Quantum Hall Effect in a Diluted Magnetic Semiconductor}
\author{C. Betthausen$^1$, P. Giudici$^1$, A. Iankilevitch$^1$, C. Preis$^1$, V. Kolkovsky$^2$, M. Wiater$^2$, G. Karczewski$^2$, B. A. Piot$^3$, J. Kunc$^3$, M. Potemski$^3$, T. Wojtowicz$^2$, and D. Weiss$^{1\star}$}

\affiliation{$^1$Department of Experimental and Applied Physics,   University of Regensburg, 93040 Regensburg, Germany.\\
$^2$Institute of Physics, Polish Academy of Sciences, 02668 Warsaw, Poland.\\
$^3$Laboratoire National des Champs Magn\'{e}tiques Intenses, CNRS-UJF-UPS-INSA, 38042 Grenoble, France.}
\date{21 Mai 2014}

\begin{abstract}

We report the observation of the fractional quantum Hall effect in the lowest Landau level of a two-dimensional electron system (2DES),  residing in the diluted magnetic semiconductor Cd$_{1-x}$Mn$_x$Te. The presence of magnetic impurities results in a giant Zeeman splitting leading to an unusual ordering of composite fermion Landau levels. In experiment, this results in an unconventional opening and closing of fractional  gaps around filling factor $\nu=3/2$ as a function of an in-plane magnetic field, i.e. of the Zeeman energy. By including the \textit{s-d}~exchange energy into the composite Landau level spectrum the opening and closing of the gap at filling factor $5/3$ can be modeled quantitatively. The widely tunable spin-splitting in a diluted magnetic 2DES provides a novel means to manipulate fractional states.

\end{abstract}
\maketitle

%Introduction

The fractional quantum Hall effect (FQHE) is a collective high-magnetic field phenomenon, originating from Coulomb repulsion of electrons confined in two dimensions. At certain fractional filling $\nu=p/q$ of the Landau levels (LLs) [$\nu=$ filling factor, $p,q$ = integers],
quantized plateaus in the Hall resistance $\rho_{xy}$ and vanishing longitudinal resistance $\rho_{xx}$ herald the presence of peculiar electron correlations \cite{Tsui1982,laughlin1983}. Here, the electrons condense into a liquid-like ground state that is separated by a gap~${\Delta}$ from the excited states. Most experiments to date have been carried out on GaAs-based systems, being still  the cleanest material system with the highest electron mobilities \cite{xia2004}. %The observation of these fragile states requires - due to the small gaps - low temperatures and high-quality samples. As a consequence, the FQHE could only be observed in few, non-magnetic materials, so far \cite{Willett1987, Boebinger1987, Bolotin2009, tsukazaki2010, Lai2004, Piot2010}. However, the main body of research has been carried out on GaAs-based systems, being still the cleanest material system with the highest electron mobilities \cite{xia2004}. Paramagnetic impurities like spin~$ 5/2$ manganese (Mn) ions, embedded into a cadmium telluride (CdTe) quantum well, constitute a diluted magnetic two-dimensional electron system (2DES)  where strong \textit{s-d}~exchange interaction between the Mn spins and the electrons prevails \cite{Furdyna1988}. This results in a giant Zeeman splitting \cite{Furdyna1988} ${E_Z}$ which is  tunable in magnitude, sign and field dependence \cite{teran2010}. %but exceeds the energy gaps~${\Delta}$ by far, thus posing the question whether the FQHE exists in diluted magnetic 2DESs at all.
When the direction of the magnetic field $B$ is tilted, the orbital LL~splitting is given by the field component~$B_{\bot}$ normal to the 2DES while the total  field strength $B$ determines the Zeeman splitting~$E_Z$. Early experiments on GaAs revealed that the $\nu = 4/3, 5/3$ and $8/5$ states behave differently upon tilting the sample \cite{Clark1989, Eisenstein1989}: While the $\nu = 4/3$ and $8/5$ states were undergoing a transition from a spin-unpolarized state to a polarized one, the $\nu= 5/3$ state is always fully spin polarized.

Although the FQHE has been reported in quite a number of different materials \cite{Willett1987, Boebinger1987, Bolotin2009, tsukazaki2010, Lai2004, Piot2010, McFarland2009}, the FQHE was never observed in a diluted magnetic semiconductor in which atoms with  magnetic moment (e.g. Mn$^{2+}$) are placed in the  2DES. Then, the localized spins in the magnetic impurities' \textit{d}-orbitals interact with the correlated electron system via the quantum mechanical \mbox{\textit{s-d}}~exchange interaction, causing giant Zeeman splitting \cite{Furdyna1988} which is  tunable in magnitude, sign and field dependence \cite{teran2010}. The constant~$\alpha N_0 \gg \Delta$ specifies the \textit{s-d}~exchange  strength and is  the largest energy scale in the system. It hence has remained unclear whether FQHE states survive in the presence of magnetic impurities.  Below we demonstrate that (i), the FQHE indeed exists in magnetic 2DESs, and
(ii), that the opening and closing of gaps in an in-plane field can be described within a modified composite fermion (CF) picture, in which the \textit{s-d}~exchange is taken into account.
%and (iii), that the Zeeman splitting of the CF-Landau levels~(CF-LLs) can be tuned from positive to negative values {(similar tuning has also been achieved by hydrostatic pressure in GaAs \cite{Leadley1997})}, thus altering the spin polarization of fractional ground states by \textit{s-d}~exchange interaction.

\begin{figure}
%fig1
\includegraphics[width=8cm]{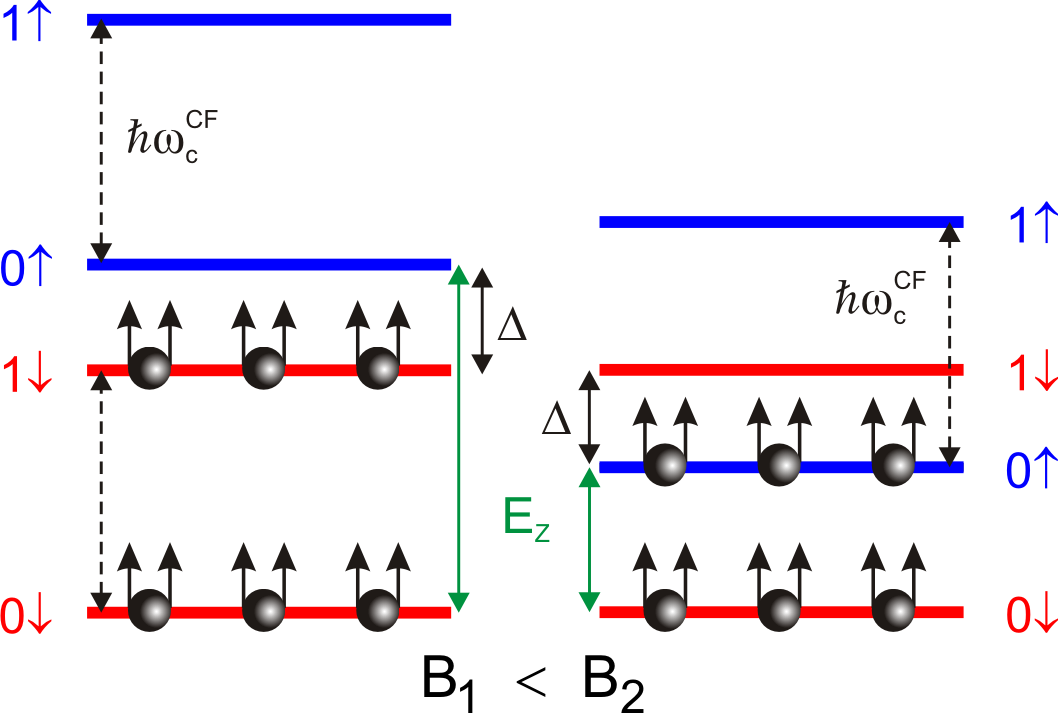}
\caption{ CF-Landau levels at filling factor $\nu=4/3$. Within the CF picture, two (spin-split) CF-LLs are occupied ($\nu^{\star} = 2$). The gap $\Delta$ is the energy difference between highest occupied and lowest unoccupied CF-LL. The CF cyclotron energy
$\hbar \omega^{\mathit{CF}}_c$,  gap $\Delta$, and the Zeeman splitting $E_Z$ are indicated. With growing in-plane $B$ (right panel) $E_Z$ decreases, causing a change of the ground state's spin polarization, different to what is expected for non-magnetic materials.
\label{fig_CF_LLs}}
\end{figure}

Let us first recall the CF-model which maps the FQHE onto the integer quantum Hall effect~(IQHE) by introducing new particles,  composite fermions, each composed of an electron and an even number (here: 2) of flux quanta \cite{Jain1989}. %These CFs are only weakly interacting and the FQHE can be understood as the IQHE of CFs.
Between $1 < \nu < 2$ the effective magnetic field for CFs vanishes at $\nu=3/2$ while they encounter an effective magnetic field $B_{\mathit{CF}}$ %=3(B_\bot-B_{3/2})$
 away from this filling \cite{Du1995}. %where $B_\bot$ is the magnetic field normal to the 2DES and $B_{3/2}$ is the normal component of the field at $\nu=3/2$.
In the vicinity of $\nu=3/2$ the CF~filling factor~$\nu^\star$ {for composite fermions of holes} is related to the one of electrons via $\nu= 2-\nu^\star /(2 \nu^\star \pm 1)$ where `$\pm$' relates to CF~filling factors at positive and negative effective  field~$B_\mathit{CF}$ \cite{Du1995}. Within this picture the fractions at $\nu=5/3$ and $4/3$ correspond to a filling of CF-LLs of $\nu^{\star}=1$ and $2$, respectively. Oscillations of $\rho_{xx}$ and steps in $\rho_{xy}$  around $\nu = 3/2$ then reflect Shubnikov-de Haas~(SdH) oscillations and the IQHE of CFs  which occupy CF-LLs separated by energy gaps $\Delta = \hbar \omega^{\mathit{CF}}_c$ with $ \omega^{\mathit{CF}}_c=eB_{\mathit{CF}}/m_{\mathit{CF}}$ ($m_{\mathit{CF}} =$ composite fermion mass). Fig.~\ref{fig_CF_LLs} illustrates CF-LLs in Cd$_{1-x}$Mn$_x$Te at fixed $B_\bot$ for two values of the in-plane magnetic field.

\begin{figure}
%fig2
\includegraphics[width=11cm]{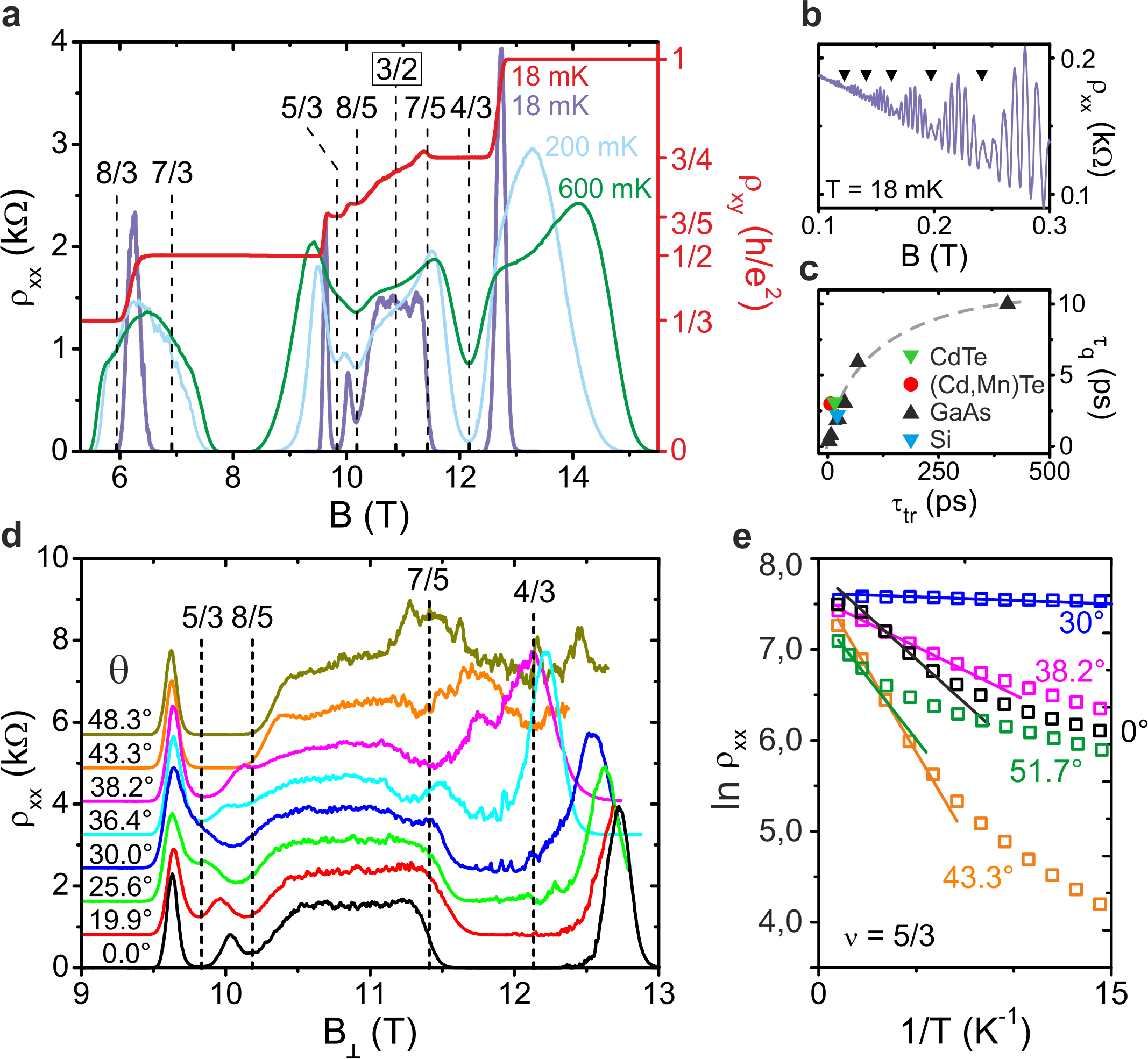}
\caption{(a)~ $\rho_{xy}$ and  $\rho_{xx}$ at various temperatures after illumination. The $B-$field is applied perpendicular to the 2DES, filling factors are indicated. (b)~Low field $\rho_{xx}$ data at $T=18$~mK. The distinct beating pattern  stems from the relative shift of spin-up and spin-down LLs due to the giant Zeeman splitting and is used to determine the Mn-concentration $x$. (c)~Quantum scattering time $\tau_q$, extracted from the low field damping of SdH oscillations vs. momentum relaxation time $\tau_{tr}$  for different systems. Data for Si taken from  \cite{Lai2005}, for GaAs from  \cite{Coleridge1991,Zumbuhl2012} and for CdTe from  \cite{Piot2010}. $\tau_{tr}$ and $\tau_q$ of our (Cd,Mn)Te QW fit nicely in the evolution of these parameters. The dashed line is a guide to the eye.
(d)~Angular dependence of $\rho_{xx}$ as  function  $B_\bot$ in the vicinity of $\nu=3/2$ at $T=25$~mK. With increasing tilt angle $\theta$ the in-plane component of the magnetic field $B_\parallel = B \sin\theta$ increases, thus changing $E_Z$. Here, $B$ is the total applied field strength.  Traces are shifted for clarity.
(e)~ $\rho_{xx}$  vs. $1/T$ on a semi-logarithmic scale for $\nu=5/3$ at various tilt angles $\theta$. Activation energies are extracted  from {Arrhenius plots (solid lines)}.
\label{FQHE_CdMnTe} }
\end{figure}

Our samples consist of $30$~nm wide diluted magnetic Cd$_{1-x}$Mn$_x$Te quantum wells sandwiched between Cd$_{0.71}$Mg$_{0.29}$Te barriers; the wells are single-sided modulation doped with iodine. Several Cd$_{1-x}$Mn$_x$Te quantum structures from two different wafers (Mn concentrations of $x=0.24$\% and $0.30$\%) have been studied. %All show similar results.
Here, the Mn concentration is obtained from the beating pattern of SdH~oscillations at low $B$ \cite{Teran2002} (Fig.~\ref{FQHE_CdMnTe}b).       % which stems from superposition of contributions from spin-up and spin-down Landau levels shifted by the Zeeman energy~$E_Z$.
Below we focus on data obtained  from one sample with especially well developed FQHE states, having $x=0.24$\%, a low-temperature mobility and carrier density of $\mu=115\,000$~cm$^2$/Vs and $n_s=3.95\cdot10^{11}$~cm$^{-2}$, respectively, after illumination with a yellow LED. Illumination turned out to be crucial to observe fractional states as it increases the quantum scattering time~$\tau_q$ by a factor of 5 to a value of $\sim 3$~ps. It is thus comparable to the values observed in GaAs and Si based 2DESs  (see Fig.~\ref{FQHE_CdMnTe}c) \cite{footnote1}.  Measurements  were done on rectangular samples of size 1.5~$\times$~3~mm$^2$  with alloyed indium contacts  in a $^3$He/$^4$He dilution refrigerator  with in-situ sample rotation stage.

%The samples are cooled down in a toploading $^3$He/$^4$He dilution refrigerator (accessible temperature range 15~mK to 1~K) equipped with a 19~T superconducting magnet and an in-situ sample rotation stage. Both the sample and a calibrated $RuO_2$ resistor thermometer are immersed in the liquid of the mixture. Our thermometer shows only a weak magnetic field dependence; temperatures measured at high fields deviate by at most $10\%$ from the values at $B=0$. High-frequency radiation that may heat up the electronic system is blocked by $\pi$-filters attached to all wiring. Resistances are probed either in the dark or after illumination with a yellow LED using standard low-frequency lock-in technique with low excitation currents of typically 10~nA to avoid heating effects.
%\textcolor{red}{\sout{The resulting electron densities before and after illumination are $1.86\cdot10^{11}$~cm$^{-2}$ and $3.95\cdot10^{11}$~cm$^{-2}$, respectively with the mobility peaking at $115\,000$~cm$^2$/Vs in the illuminated case.}}

%Experiments
Fig.~\ref{FQHE_CdMnTe}a depicts $\rho_{xy}$ and $\rho_{xx}$
of our Cd$_{1-x}$Mn$_x$Te quantum well~(QW) device
%versus $B_\bot$
in perpendicular magnetic field for different temperatures T.
% with a yellow LED.
%The specimen features an electron density of $3.95\cdot10^{11}$~cm$^{-2}$, its mobility peaks at $115\,000$~cm$^2$/Vs after
%illumination.
Pronounced minima corresponding to  $\nu=5/3$, $8/5$ and $4/3$ emerge in the lowest LL at low $T$  around $\nu=3/2$. No minimum is observed at $\nu=7/5$. %Similar to the case of CdTe \cite{Piot2010} weak FQH minima ($\nu=7/3$ and $8/3$) emerge around $\nu=5/2$ at temperatures above $T\approx 450$~mK. As these states are masked by the IQHE at lowest temperatures and further display an analogous characteristics as their counterparts in CdTe we will not discuss details here.
%Note that the presence of manganese in the QW is evidenced by the appearance of a distinct beating pattern with nodes \cite{Teran2002} at low magnetic fields (see inset of Fig.~\ref{FQHE_CdMnTe}a).
A set of $\rho_{xx}$ data, taken at $T=25$~mK for different tilt angles, is shown in Fig.~\ref{FQHE_CdMnTe}d. %depicting a set of angular dependent magnetoresistance traces recorded at a temperature of $T=25$~mK.
Tilting the sample by an angle~$\theta$  yields the perpendicular magnetic field as $B_\bot=B\cdot\cos\theta$. A rather complex angular dependence appears in the $\rho_{xx}(\theta)$ traces, i.e. minima disappear and reappear. The $\nu=5/3$ minimum for instance starts - in clear contrast to non-magnetic 2DESs - to weaken continuously as soon as the sample is tilted away from the initial $\theta=0^\circ$ position, completely vanishes at about $\theta=30^{\circ}$ and reemerges upon further tilt.  This is in stark contrast to the angular dependence of the $5/3$ minimum in GaAs or CdTe where no weakening of the gap was found. Notably, the $\nu=7/5$ minimum, absent in perpendicular magnetic field, appears around $\theta=38^{\circ}$.

\begin{figure}
%fig3
\includegraphics[width=7 cm]{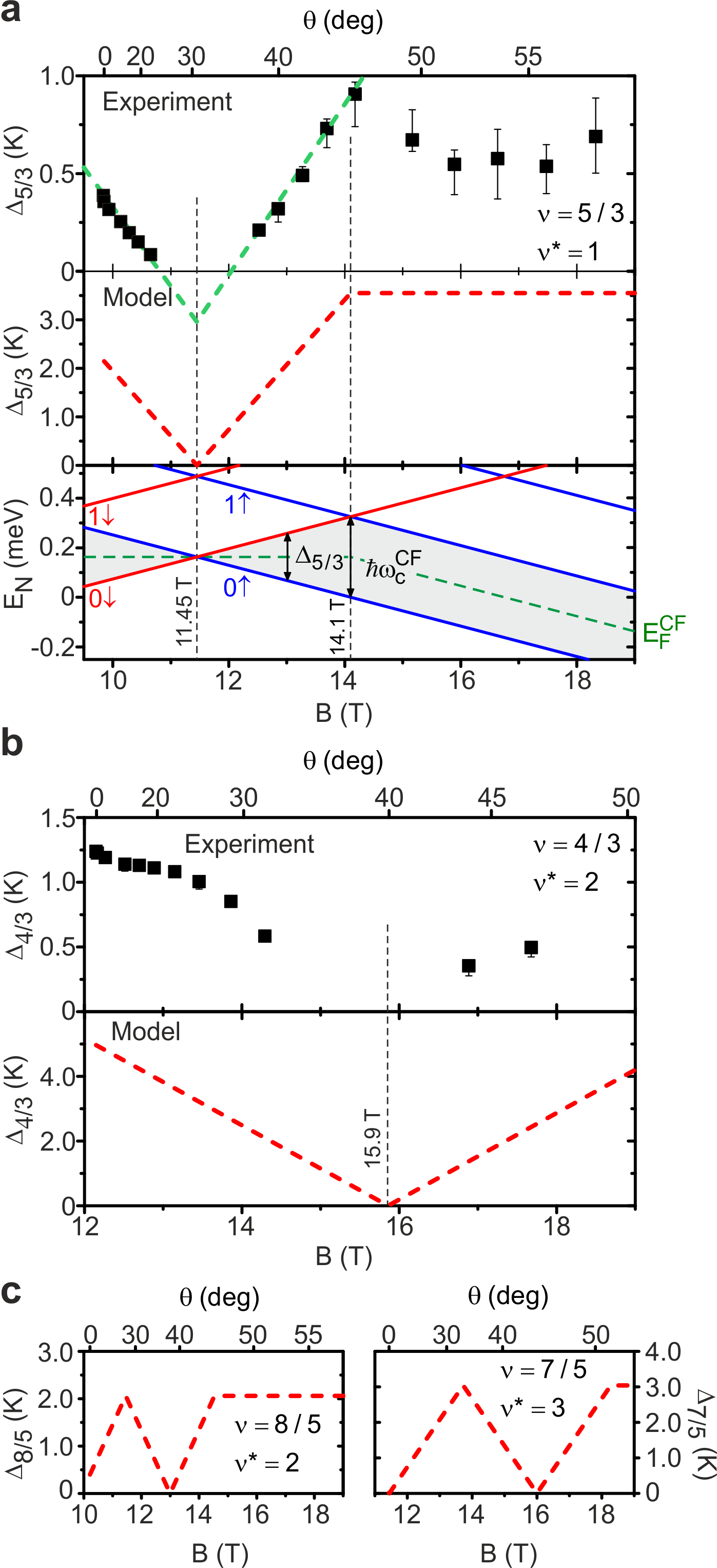}
\caption{ (a)~(Top panel) Activation gaps $\Delta_{5/3}(B)$, taken at fixed $B_\bot=9.85$~T as function of total field  $B$ (tilt angle $\theta$). The $B-$field position at which $\Delta_{5/3}$ closes was obtained by extrapolation (green dashed line). (Middle panel)~Modeled gap $\Delta_{5/3}(B)$ as deduced from the CF-LL scheme of the $\nu=5/3$ ($\nu^\star =1$) state shown in the bottom panel. (Bottom panel)~CF-LLs (Eq.~\ref{eqn_CF_LLs}) as function of total field $B$ for fixed $\hbar \omega^{\mathit{CF}}_c$, i.e. fixed $B_\bot$. The vanishing gap at $B=$~11.45~T is assigned to  $N=0_\downarrow$ and $0_\uparrow$ CF-LL crossings while the maximum $\Delta_{5/3}$ at $B=$~14.1~T corresponds to the cyclotron gap $\hbar \omega^{\mathit{CF}}_c$.  %Parameters $\hbar \omega^{\mathit{CF}}_c$ and $g^\star$ are adjusted to reproduce minima and maxima of $\Delta_{5/3}(B)$. 
(b)~Activation gap $\Delta_{4/3}$ (top) and corresponding model $\Delta_{4/3}$ (bottom). (c)~Calculated evolution of the gaps $\Delta_{8/5}(B)$ and $\Delta_{7/5}(B)$ for the parameters given in the text.}
\label{gaps_vs_tilt}
\end{figure}

To quantify our observations we performed angular dependent activation energy measurements,  shown in Fig.~\ref{FQHE_CdMnTe}e for the $\nu=5/3$ state. There, $\rho_{xx}$ at $\nu=5/3$ is recorded as a function of temperature for various $\theta$                                                               (i.e. for the same~$B_\bot$).
Activation gaps~$\Delta_{5/3}$ are then obtained from Arrhenius plots.
%by fitting the $\rho_{xx}(T)$~data with a model that considers disorder-broadened LLs of gaussian shape \cite{Usher1990}. %Thereby we account for deviations from simple thermally activated transport, arising if LL broadening is no longer negligible in comparison to $\Delta_\nu$.
Corresponding data of  $\Delta_{5/3}$ are shown in
Fig.~\ref{gaps_vs_tilt}a (top panel) as function of the total field $B$:
%Let us for example focus on the $\nu=5/3$~state, analysed in detail in Fig.~\ref{gap_v53}:
The gap~$\Delta_{5/3}$ starts to close when increasing $B$ by tilting the sample, vanishes  at around 11.45~T ($\theta\approx 30^\circ$),
opens again and reaches a maximum value at 14.1~T ($\theta\approx 45^\circ$). The magnetic field at which the gap vanishes is obtained by extrapolating the data points left and right of the minimum.  This gives $B=11.45$\,T ($\theta\approx 30^\circ$) \cite{footnote2}. 
This behavior, i.e. closing and opening of the gap is in line with the observed disappearance and reappearance of the $\nu=5/3$ minimum presented in Fig.~\ref{FQHE_CdMnTe}d.

%Modelling
To model the $\Delta_\nu (B)$ characteristics we expand the CF model \cite{Jain1989} to cover diluted magnetic semiconductors (DMSs).
%The CF~approach maps the FQHE of electrons onto the IQHE of non-interacting composite fermionic particles. Around $\nu=3/2$ these particles encounter an effective magnetic field \cite{Du1995} $B_{\mathit{CF}}=3(B_\perp-B_{3/2})$, where $B_{3/2}$ is the perpendicular field component at $\nu=3/2$. They are consequently quantized into CF-LLs, separated by an energy of $\hbar \omega^{\mathit{CF}}_c$, with $\omega^{\mathit{CF}}_c=e |B_{\mathit{CF}}|/m_{\mathit{CF}}$ and the CF effective mass $m_{\mathit{CF}}$.
%Because the CF cyclotron gap $\hbar \omega^{\mathit{CF}}_c$ arises due to Coulomb interaction between the electrons, we expect it to be proportional to $\sqrt{B_\bot}$. Consequently, we assume the CF~mass to be given by $m_{\mathit{CF}}=C_\nu m_0\sqrt{B_\bot}$, where $m_0$ is the bare electron mass and $C_\nu$ a constant being material and filling factor dependent \cite{Park1998}.
%In the vicinity of $\nu=3/2$ the CF~filling factor~$\nu_{\mathit{CF}}$ is related to the one of electrons via $\nu=(3\nu_{\mathit{CF}}\pm 2)/(2 \nu_{\mathit{CF}} \pm 1)$.
To account for fractional states in a DMS  we put forward a CF-LL fan chart modified by \textit{s-d}~exchange interaction,
%propose the CF~energy spectrum to be given by
\begin{eqnarray} \label{eqn_CF_LLs}
E_{N,\uparrow \downarrow} &=& \left(N+\frac{1}{2}\right)\hbar \omega^{\mathit{CF}}_c \pm
\frac{1}{2} \left[g^\star \mu_B B +\alpha N_0 x S\mathcal{B}_S%S\left(\frac{Sg\mu_B B}{k_B(T+T_{AF})} \right)
\right].
\end{eqnarray}
The first term represents the cyclotron energy of a CF in the $N^\mathrm{th}$~CF-LL, the second one describes Zeeman splitting in the presence of magnetic impurities \cite{Furdyna1988}. It has two contributions: the first part, linear in $B$, is the conventional Zeeman term with $g^\star$, viewed as the $g$-factor of CFs. The second part is due to exchange  between CFs and manganese spins with $S=5/2$, described by the \textit{s-d}~exchange constant~$\alpha N_0$ and the Brillouin function~$\mathcal{B}_S$.%, accounting for the buildup of Mn magnetisation with increasing $B$ and decreasing $T$.
% in which $S=5/2$ and $g=2$. Antiferromagnetic interactions between manganese spins in Cd$_{1-x}$Mn$_x$Te lead to a reduction of their overall magnetisation \cite{Furdyna1988} which is taken into account by the functions $x_{\mathit{eff}}(x)<x$ and $T_{AF}(x)>0$; here $x_{\mathit{eff}}=0.0023(2)$ and $T_{AF}=84(7)$~mK.
%in the studied sample
Since $\mathcal{B}_S$ saturates at $B$-fields around 1~T at low $T$, the exchange contribution to spin splitting is constant for fields above 10~T, applied here. As the $g$-factor of electrons %(and hence of CFs, see below)
in CdTe is negative while the exchange contribution is positive \cite{Furdyna1988,ganichev2009},
 $E_Z$ decreases above 1~T and eventually vanishes if both terms are of equal strength, causing an unusual ordering of CF-LLs. The transition from a polarized fractional ground state to an unpolarized one with increasing in-plane field (sketched in Fig. 1) is one example of such an unusual ordering.

To retrace the $\Delta_\nu (B)$ characteristics we first adjust the
CF-LL schemes to the experimental data at 5/3 filling. As illustration, we show in the bottom panel of Fig.~\ref{gaps_vs_tilt}a the CF-LL scheme of the $\nu=5/3$ ($\nu^\star=1$) state as a function of $B$. The gap~$\Delta_{5/3}$ (gray shaded region) is  given by the energy difference between the CF-LLs above and below the Fermi energy~$E^{CF}_F$. The bottom line here is that a vanishing gap at a specific $B$-value coincides with the crossing of  spin-up and -down, N=0, CF-LLs  while a maximum gap corresponds to the  separation between $N=0$ and $N=1$ CF-LLs with the same spin. By properly assigning $g^\star$, $\hbar \omega^{\mathit{CF}}_c$ and $\alpha N_0$, the magnetic field positions (for fixed $B_\bot$) at which a gap $\Delta_\nu$ opens or closes can be described quantitatively correct. %In an ideal system the three parameters affect the CF-LL spectrum in the following way: The CF~Landau level spacing $\hbar \omega^{\mathit{CF}}_c$ ($B_\bot$ is fixed for each $\nu$) defines the maximum gap; the CF~$g$-factor~$g^\star$ is directly related to the slopes of the $\Delta_\nu(B)$~traces via $|\partial\Delta_\nu /\partial B|=|g^\star \mu_B|$ and all three parameters together determine the crossing points of different CF-LLs on the $B$-axis. However, in the presence of disorder and LL mixing the magnitude of the gap, which is entirely due to electron-electron interactions, is reduced, and the experimental slopes $\partial\Delta_\nu /\partial B$ do not represent a reliable physical quantity anymore.

To compare model and experiment we assume that $\alpha N_0 = 220$\,meV is the same for electrons and CFs. In case of $\nu = 5/3$  the vanishing gap can be ascribed to the crossing of the $0,\!\uparrow$ and $0,\!\downarrow$ CF-LLs, occurring at  vanishing spin splitting, so that $E_Z=g^\star \mu_B B +\alpha N_0 x S=0$. Using that and the value at which the gap vanishes, $B=11.45$\,T, $x=0.24$\% and $S=5/2$ we obtain $g^\star = -1.99$. This value deviates by about 16\% from  the $g$-factor of electrons in CdTe, $g=-1.67$; similar matching between $g$-factors of electrons and CFs has been seen in experiments on GaAs heterostructures   \cite{Du1995}. Having fixed $g^\star = -1.99$ and $\alpha N_0 = 220$\,meV, we now use the "coincidence method" to determine  $\hbar \omega^{\mathit{CF}}_c$. The coincidence of the  $0,\!\uparrow$ and $1,\!\downarrow$ level occurs when the  gap $\Delta_{5/3}$ reaches its maximal value at $B=14.1$\,T (see Fig.~\ref{gaps_vs_tilt}a). Then, we have that $E_Z(B=14.1 \rm{T})=\hbar \omega^{\mathit{CF}}_c$ %at $B_\mathit{CF}^{(5/3)}$, (where $B_\mathit{CF}^{(5/3)}$ is the effective CF field at $\nu=5/3, B_{\perp}=9.85$\,T)
and obtain  $\hbar \omega^{\mathit{CF}}_c=3.54$\,K $(m_\mathit{CF}=1.25 m_e)$ at  $\nu=5/3$. The calculated evolution of $\Delta_{5/3}(B)$ (middle panel of Fig.~\ref{gaps_vs_tilt}a) reproduces the experimental data well for tilt angles below $\theta \sim 45^\circ$. With increasing in-plane field (tilt angle), however, the model describes the data less perfect. In Fig.~3a (middle panel) the model predicts a constant gap above 14~T while the data  deviate. This is to some extent due to the larger error in extracting the gap; however, a reduction of the gap is also expected from the coupling of the growing in-plane field to the orbital motion of the electrons in our 30~nm wide QW \cite{piot2009}. The gap $\hbar \omega^{\mathit{CF}}_c$, obtained by the "coincidence method" is for $\nu = 5/3$ (and also for $\nu = 4/3$, see below)
by a factor of $\sim 3$ larger than the activation gaps.  Generally, gaps are overestimated by theory when disorder, LL mixing and  finite thickness correction are neglected \cite{dean2008,wojs2006,zhang1986, morf2002}. The difference in gap size %between the top panel and the one below in Figs.~\ref{gaps_vs_tilt}a,b
 reflects the different experimental techniques used to extract  $\hbar \omega^{\mathit{CF}}_c$: The fitting of the CF-LL spectrum (red dashed lines in Fig.~\ref{gaps_vs_tilt}a,b), obtained from the coincidence of different CF-LLs, is less affected by disorder. The cyclotron gap extracted from activated transport (top panels), in contrast, is strongly affected by disorder broadening and hence  smaller \cite{leadley1998}. %A similar discrepancy has been observed for gaps between spin-split LLs in GaAs heterojunctions .%for which the activation gap is significantly smaller than the intrinsic gap, obtained, e.g., from the coincidence between different spin-split LLs

Assuming that the usual CF-LL spectrum gets modified by the exchange energy $\alpha N_0 = 220$\,meV enables us to model the gap evolution $\Delta_{5/3}(B)$ quite reasonably. Below we check whether the modified CF-LL spectrum,  extracted above  is consistent with the  observations made at other filling.
 For that we assume that  $\alpha N_0 = 220$\,meV and $g^\star = -1.99$ are independent of $\nu$. The latter assumption is justified by previous work on GaAs where the CF g-factor was found to be independent of $\nu$ and close to the one of electrons  \cite{Du1995,Du1997}. In other words, we assume that the Zeeman gap closes at the same $B=11.45$\,T for all $\nu$. For $\nu = 4/3$ we extract activation gaps of order 1\,K for small $\theta$ (top panel, Fig.~\ref{gaps_vs_tilt}b) which decrease for higher tilt angles. For $\theta$ between $~35^\circ$ and $~42^\circ$ maxima emerge at $\nu=4/3$ so that we refrain from extracting activation energies in the region where the gap closes. At still higher tilt angles the gap reappears and the data allow to extract some activation gaps although the traces in Fig.~\ref{FQHE_CdMnTe}d display some noise. The overall dependence of $\Delta_{4/3}(B)$, modeled with the same parameters $\alpha N_0$ and  $g^\star$ as above (Fig.~\ref{gaps_vs_tilt}b, bottom panel) agrees well with the  data plotted in the top panel. This time, however, we assume that $\hbar \omega^{\mathit{CF}}_c(\nu=4/3)=6$\,K $(m_\mathit{CF}=0.94m_e)$.
 %At variance with the previous case of $\nu=5/3$, it is now, for $\nu=4/3$   $(\nu^\star=2)$,
 Here, the $B$-field at which $\Delta_{4/3}$ vanishes  defines the CF cyclotron gap. We note, though, that the  data points do not allow to determine the $B$-value at which the gap closes accurately. We have chosen  $B=15.9$\,T; shifting this point by a Tesla towards smaller fields reduces $\hbar \omega^{\mathit{CF}}_c$  by about 20\% and leads a corresponding rescaling of the energy axis. Summarizing this point we note that the evolution of $\Delta_{4/3}(B)$ is consistent withthe CF-LL scheme deduced above.%where spin-up and spin-down LLs are additionally shifted by the exchange energy.

Finally, let us now turn to the gaps at  $\nu=7/5$ and $8/5$. While we observed gaps for  $\nu=5/3$ and $4/3$ in different samples, fractional states at  $7/5$ and $8/5$ filling were only observed in the sample discussed here. Because the extraction of gaps might be disputable since for most tilt angles no clear $\rho_{xx}$ minima arise we only show modeled $\Delta_{7/5}(B)$ and $\Delta_{8/5}(B)$ traces to illustrate that their evolution is, within experimental accuracy, consistent with the  resistivity as function of $\theta$. Here we assume that maxima in the resistivity, measured as a function of $\theta$ (i.e. the resistivity along the dashed lines in Fig.~\ref{FQHE_CdMnTe}d) correspond to vanishing CF-LL gaps. Using, as before, $\alpha N_0 = 220$~meV and $g^\star = -1.99$ we can obtain a vanishing gap at $\sim 13$~T for filling factor $8/5$ in Fig.~\ref{gaps_vs_tilt}c (left panel). This agrees with a maximum of $\rho_{xx}$ at  $\theta \sim 38^\circ$ in Fig.~\ref{FQHE_CdMnTe}d. For  $\nu = 7/5$ maxima in $\rho_{xx}$ at $\theta \sim 8^\circ$ (not shown) and  $\sim 48^\circ$ correspond to vanishing gaps at about $\sim 11.5$~T and $17$~T in Fig.~\ref{gaps_vs_tilt}c (left panel). This agrees reasonably well with the  calculated traces.

%Conclusions
In summary we note that without taking $\alpha N_0$ in Eq.~(1) into account the angular dependence of $\Delta_{\nu}(B)$ cannot be reproduced. Especially the closing and opening of the $\nu = 5/3$ gap, not observed in any other material, and corresponding to a change in spin polarization from $0,\!\downarrow$ to $0,\!\uparrow$ at $B=11.45$\,T, highlights the impact of  exchange splitting on the spin-polarization of the CF ground states. %Thus Cd$_{1-x}$Mn$_x$Te presents an interesting system for the investigation of spin-related effects in the FQHE regime since it offers the possibility to tune the spin splitting~$E_Z$ to zero at nearly any desired $B$-field  via changing the in-plane field or the Mn content of the sample.

%Our results demonstrate the emergence of FQHE states in a Cd$_{1-x}$Mn$_x$Te QW despite the presence of magnetic impurities. The latter - by giving rise to \textit{s-d}~exchange interaction - strongly affect the angular evolution of activation energy gaps which we explain within an extended composite fermion model.
%Thus Cd$_{1-x}$Mn$_x$Te presents an interesting system for the investigation of spin-related effects in the FQHE regime since it offers the possibility to tune spin splitting~$E_Z$ to zero at nearly any desired magnetic field value (via changing the in-plane field or the manganese content of the sample). %The tunability of the Zeeman splitting might for instance help to answer the open question whether the $\nu=5/2$ state (possibly emerging in Cd$_{1-x}$Mn$_x$Te with increasing sample quality) could form skyrmions or spin topological defects at $E_Z=0$.
%Thus, by increasing the quality of Cd$_{1-x}$Mn$_x$Te QW similar as in the GaAs case (and there is no principle obstacle which impedes this) diluted magnetic 2DESs can become a new model system for further investigation of states in the FQHE regime because of the additional degree of freedom, the exchange splitting, which can be tuned in a wide range.

We thank A.~W\'{o}js for fruitful discussions. This work was supported by the German Science Foundation (DFG) via Grants GI539, WE2476 and by the European Commission within the programme "Translational Access" (Contract  228043-EuroMagNET II). The research in Poland was partially supported by the National Science Centre (Poland) under grant DEC-2012/06/A/ST3/00247.
\clearpage

%\bibliography{Literature}

\begin{thebibliography}{10}
\bibitem{footnote2}
The negative gap at the crossing point can be ascribed to the finite CF Landau level width.
\end{thebibliography}

\end{document}